\documentclass{PoS}
\usepackage{cite}

\title{
\vspace*{2mm}
{\tiny DESY 10--103 \hfill SFB/CPP-10-66}\\ 
Update of the NNLO PDFs in the 3-, 4-, and 5-flavour schemes}

\ShortTitle{Update of the NNLO PDFs in the 3- and 5-flavour scheme}

\author{\speaker{Sergey Alekhin}\\
        Deutsches Elektronen-Synchrotron, DESY, Platanenallee 6,
        D-15738 Zeuthen, Germany  \& \\
        Institute for High Energy Physics, IHEP, Pobeda 1, 142281 Protvino, 
        Russia\\ 
        E-mail: \email{sergey.alekhin@ihep.ru}}

\author{Johannes Bl\"umlein\\
        Deutsches Elektronen-Synchrotron, DESY, Platanenallee 6,
        D-15738 Zeuthen, Germany\\
        E-mail: \email{Johannes.Bluemlein@desy.de}}

\author{Sven-Olaf Moch\\
        Deutsches Elektronen-Synchrotron, DESY, Platanenallee 6,
        D-15738 Zeuthen, Germany\\
        E-mail: \email{sven-olaf.moch@desy.de}}

\abstract{We report on an update of the next-to-next-to-leading order 
(NNLO) ABKM09 
parton distributions functions. They are obtained
with the use of the combined 
HERA collider Run I inclusive deep-inelastic scattering (DIS) data and the 
partial NNLO corrections to the heavy quark electro-production
taken into account. The value of the strong couplig constant    
$\alpha^{\rm NNLO}_{\rm s}(M_{\rm Z})=0.1147(12)$ is obtained. 
The standard candle cross sections for the Tevatron 
collider and the LHC estimated with the updated PDFs are provided.
}

\FullConference{XVIII International Workshop on Deep-Inelastic Scattering and Related Subjects\\
                 April 19 -23, 2010\\
                 Convitto della Calza, Firenze, Italy}

\begin{document}

The parton distribution functions (PDFs) are an essential ingredient for 
hadron collider phenomenology. With increasing accuracy of the data and 
expansion of their kinematics, particularly due to the start-up of the LHC, 
further validation of the PDFs is required. This implies 
both employment of new, more accurate data sets 
and theoretical improvement in the data interpretation. These two issues
are usually related since higher accuracy of the data often demands 
to improve the theoretical accuracy as well. Particularly precise PDFs 
are necessary for the calculation of the 
standard candle cross sections, as for 
Higgs-, top quark-, 
and $W/Z$ hadro-production. To provide a theoretical accuracy
of O(1\%) for those processes one has to take into 
account the next-to-next-to-leading order (NNLO) QCD corrections. Therefore 
the PDFs extracted in the NNLO QCD approximation are required.

In the following we report an update of the NNLO PDF 
set of Ref.~\cite{ABKM}, which are 
extracted from the data sets including the deep-inelastic-scattering 
(DIS) data obtained by the HERA collider experiments. 
The latter give an important constraint on the 
PDF kinematics relevant for the interpretation of the LHC data 
to be collected in the first run. The inclusive DIS data obtained 
by the H1 and ZEUS collaborations in the Run I 
of the HERA collider~\cite{H1ZEUS}
were recently merged into one combined data set~\cite{HERA}. This 
set replaces the separate H1 and ZEUS data sets used
in the analysis of Ref.~\cite{ABKM}. In order to illuminate the 
trend of the new data set with respect to the previous version of the fit 
we first perform a model independent analysis of the combined HERA data
on the inclusive structure function $F_2$. 
To obtain the values of $F_2$ we correct
the cross section values of Ref.~\cite{HERA} for the contribution due to the 
longitudinal structure function $F_{\rm L}$, which is calculated making use of 
the NNLO PDFs of Ref.~\cite{ABKM}. The data points obtained
are separated by bins in the Bjorken variable $x$. For each bin 
a second-order polynomial in $\ln(Q^2/3~{\rm GeV}^2$), 
where $Q^2$ denotes the momentum transferred, is fitted to the data.
In such a way we obtain the constant term, slope, and curvature 
of $F_2$ with respect to $\ln (Q^2$), depending on $x$.  
To minimize the bias in the determination of the $F_2$-slope, which often 
serves as an important
indicator of the QCD dynamics, we use  only the data 
in the range of  $2~{\rm GeV}^2<Q^2<100~{\rm GeV}^2$ for this fit. 
\begin{figure}[h]
\includegraphics[width=\textwidth,,height=5.5cm]{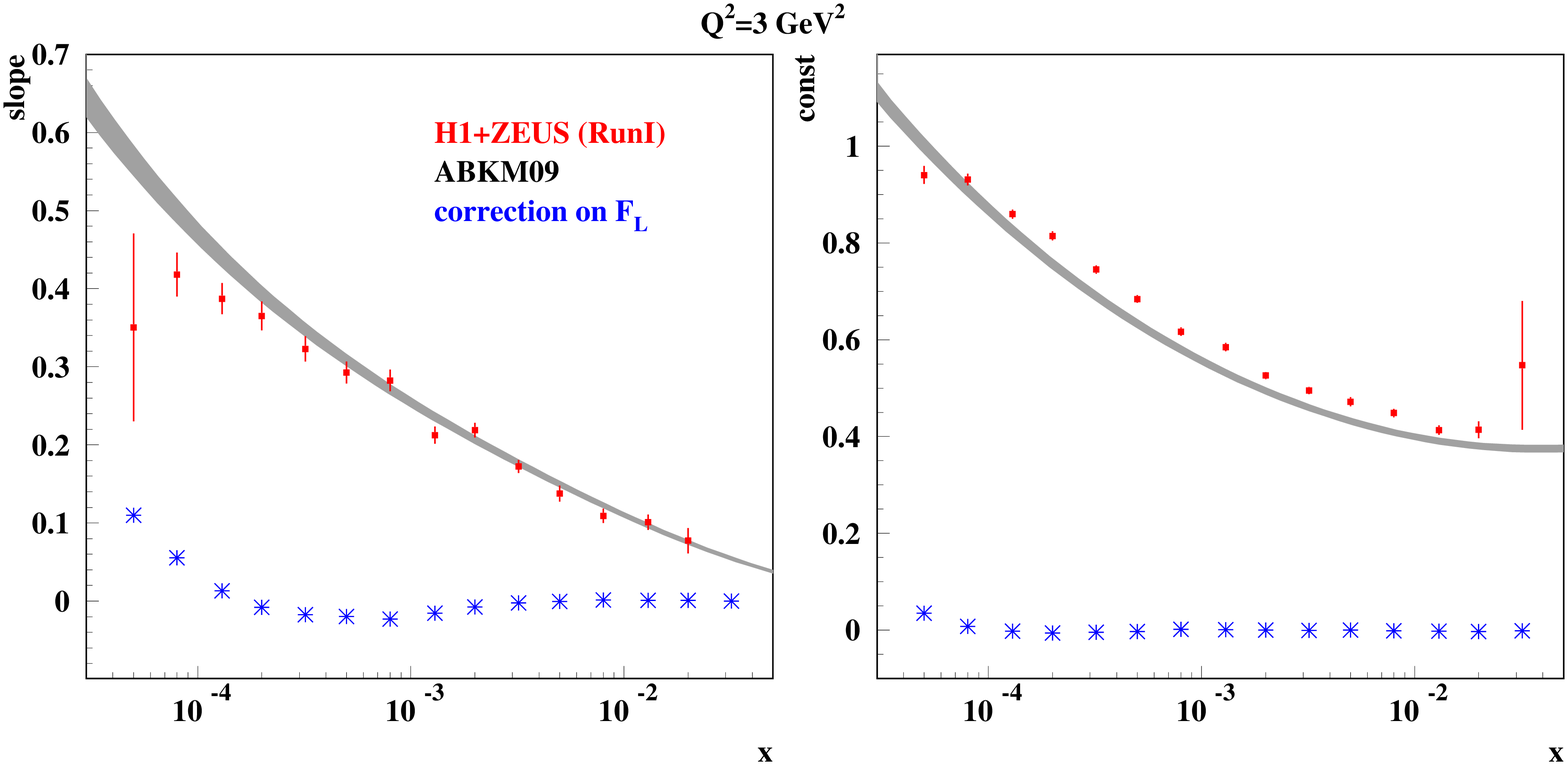}
\caption{The slope on $\ln (Q^2)$ 
(left panel) and the constant term (right panel)
obtained in the model-independent analysis of the combined HERA Run I data 
for the inclusive structure function $F_2$ in comparison to the predictions of 
ABKM09 fit~\cite{ABKM}. The impact of the correction on the 
contribution of longitudinal 
structure function $F_{\rm L}$ employed to extract the
 values of $F_2$ from the data on the 
cross sections is given by the stars.} 
\label{fig:hera}
\end{figure}
With the systematic error correlations taken into account
the value of $\chi^2/NDP$ obtained is $153/134=1.14$. 
Thus the fluctuations in data are somewhat bigger than the uncertainties
quoted in Ref.~\cite{HERA}.
On the other hand, in the variant of the fit  
with the systematic errors combined with the statistical ones in quadrature
the value of $\chi^2/NDP=123/134=0.9$ is obtained.
Therefore in this case the fluctuations in the data 
are overestimated. The $x$-dependence
of the slope and the constant term in the case of the model-independent 
fit to the $e^+ p$ data sample are compared to the 
NNLO predictions based on the ABKM09 PDFs in Fig.~\ref{fig:hera}.
The $F_2$-slope is found in reasonable agreement to the predictions, while 
the constant term in general overshoots it. This happens since the 
combined HERA data go higher than the separate data of Ref.~\cite{H1ZEUS}
in general. Particularly this is the case for the H1 experiment. 
\begin{figure}[h]
\includegraphics[width=\textwidth,,height=9cm]{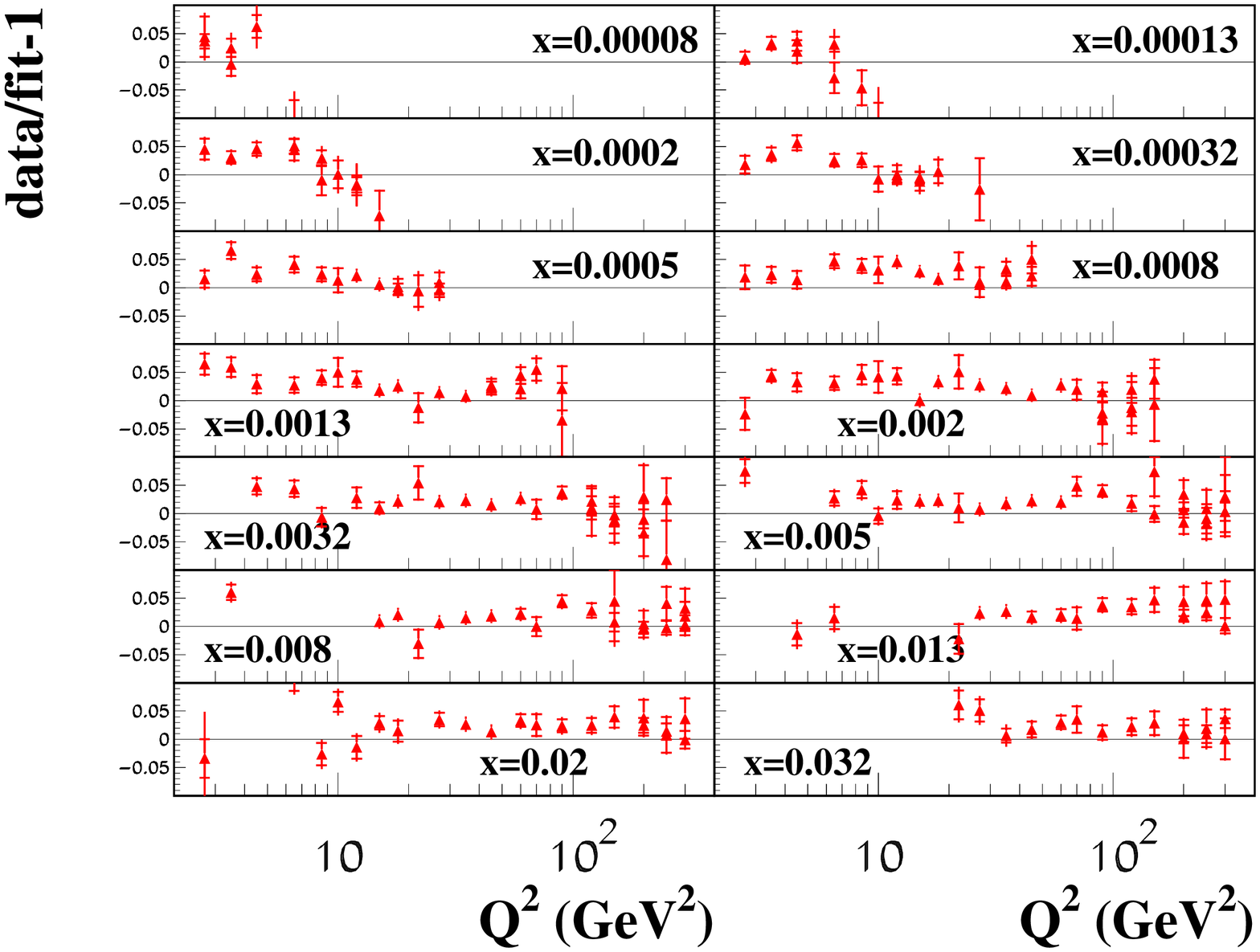}
\caption{The pulls for the low-$x$ part of the combined HERA data 
used in the updated version of our QCD fit.} 
\label{fig:pull}
\end{figure}
However, at small $x$ the trend is opposite, therefore the new data call for a 
change in the parametrization of the 
low-$x$ PDF asymptotics. To allow for a better flexibility of the 
PDFs in this region  we employ the following general shape of PDFs 
\begin{equation}
q_i(x,\mu_0)=exp\left[a\ln x(1+\beta \ln x)(1+\gamma_1 x + \gamma_2 x^2
+ \gamma_3 x^3)\right](1-x)^b
\label{eqn:pdfs}
\end{equation}
at the boundary factorization scale $\mu_0$. 
The high-$x$ asymptotics of Eq.(~\ref{eqn:pdfs}) is defined by the 
parameter $b$
and its low-$x$ asymptotics -- by two parameters, $a$ and $\beta$. 
If $\beta=0$, at low $x$ Eq.(~\ref{eqn:pdfs}) reproduces the general
Regge-motivated ansatz. The parameter $\beta$ describes possible deviations 
off this ansatz, like in Fig.~\ref{fig:hera}. 
For the valence quark distributions
such a parameter is irrelevant. We tried to fit it only for the gluon and sea 
distributions. The strange quark sea at small $x$ is poorly 
constrained by the existing data and is not sensitive to such a modification 
of the low-$x$ asymptotics. For the gluon distribution the parameter $\beta$
is also comparable to zero within uncertainties. However for the    
non-strange quark sea a small positive value of $\beta$ is preferred by 
the data, in agreement with the trend given in Fig.~\ref{fig:hera}.

A correct interpretation of the inclusive DIS data at small $x$ implies
account of the high-order QCD corrections, which are particularly large at 
this kinematic. In the fit of Ref.~\cite{ABKM} we employ the NNLO corrections 
to the DIS light parton contribution and the NLO corrections to the 
heavy-quark DIS production, which are considered in the 3-flavor
factorization scheme of $O(\alpha_{\rm s}^2)$ for the neutral current 
and of $O(\alpha_{\rm s})$ for the charged current. The NNLO corrections 
to the heavy-quark DIS production are only partially 
known Ref.~\cite{JB,ML,MITOV}. 
In particular, for the case of neutral 
current they are calculated for the heavy-quark threshold production kinematics
using the soft-gluon resummation
technique up to terms of $O(\beta_h)$~\cite{AM}, where $\beta_h$ is the 
heavy-quark velocity. 
The threshold contributions are particularly big at   
small $x$ and $Q^2$~\cite{VOGT} and brings the QCD predictions in better 
agreement with the existing HERA data on the charm semi-inclusive structure 
functions~\cite{AM}.

In the updated version of our QCD fit based on the combined HERA data
we employ the
$O(\alpha_{\rm s}^2)$ threshold corrections of Ref.~\cite{AM} in combination 
with a new, more flexible, PDF shape of Eq.(~\ref{eqn:pdfs}). 
The pulls obtained in this fit for the most accurate part of 
the HERA data with $Q^2<300~{\rm GeV}^2$  
are given in Fig.~\ref{fig:pull}.
The cut of $Q^2>2.5~{\rm GeV}^2$ is also used in our
fit and applied to the data in Fig.~\ref{fig:pull}. The value 
of $\chi^2/NDP$ for this part of the combined HERA data set is $365/202=1.2$,
which is comparable to the value obtained in our model independent fit. 
In general the data do somewhat overshoot the fit. This is 
statistically admissible, once we take into account the error correlations. 
Since the systematic errors in the data are 
significant, in this case the coherent shift of the data points by the 
value of the systematic uncertainty does not lead to a big penalty
on the value of $\chi^2$. However, this signals a tension at 
the level of $1\sigma$ between the combined 
HERA data and the fixed target DIS data
by NMC collaboration, which spread down to $x\sim 0.01$. 
\begin{figure}[h]
\includegraphics[width=\textwidth,,height=5.5cm]{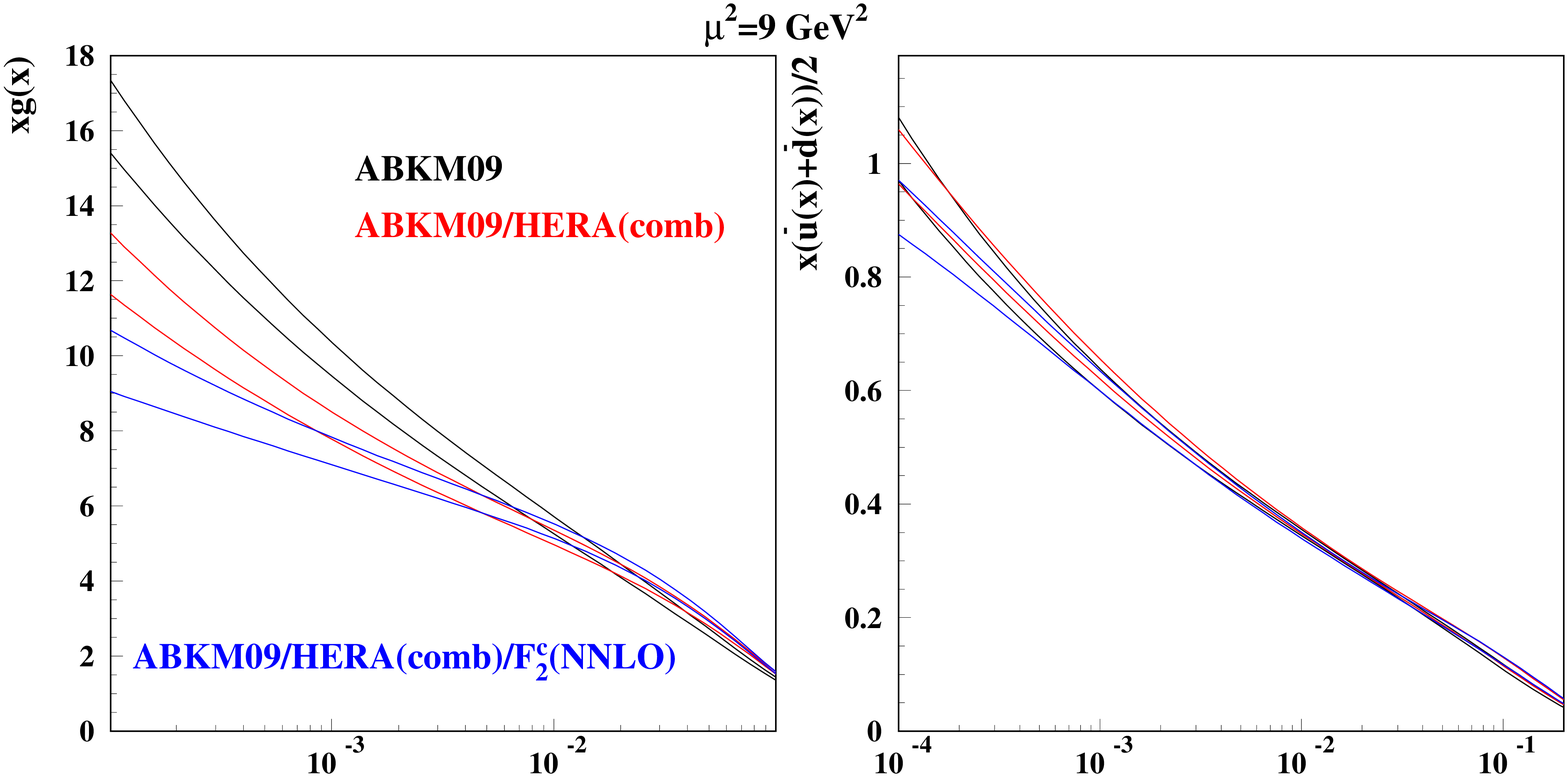}
\caption{The $1\sigma$ bands for 
the 3-flavor
 NNLO gluon distribution (left panel) and the non-strange sea distribution 
(right panel) obtained in the updated version of our fit (blue) compared to 
the same for ABKM09 fit (black) and the version of the updated fit 
without partial NNLO corrections to the heavy-quark electro-production 
taken into account (red) at the factorization scale of $9~{\rm GeV}^2$.}  
\label{fig:pdfs}
\end{figure}

The gluon and non-strange 
sea distributions at small $x$ obtained in different variants of our fit 
are compared in Fig.~\ref{fig:pdfs}. 
In the updated version the sea distribution 
at small $x$ moves upward as compared to the one of Ref.~\cite{ABKM} due to 
the impact of the update in the HERA data. The gluon distribution 
at small $x$ is correspondingly lower than that of Ref.~\cite{ABKM}, 
in particular, due to the momentum 
conservation imposed on the PDFs in the fit. The NNLO corrections 
to the heavy-quark electro-production also lead to a suppression of 
the gluon distribution at small $x$. Nonetheless, despite this 
decrease, the gluon distribution remains positive down to 
the relevant factorization scale 
of $\sim 2~{\rm GeV}^2$. The value of the strong coupling constant 
$\alpha_{\rm s}$ is fitted simultaneously with the PDFs. For the value of 
$\alpha_{\rm s}(M^2_{\rm Z})$ and in 
the 5-flavor scheme we obtain $0.1147(12)$. This is by $1\sigma$ bigger than 
the value of $0.1135(14)$, which was obtained in the analysis of 
Ref.~\cite{ABKM}, mainly due to the impact of the update in the HERA data. 

The cross sections of the standard candle processes at 
the energies of the Tevatron collider and at LHC
calculated with the use of updated PDFs
are given in Table~\ref{tab:cs}. The same cross sections 
obtained for the previous version of our PDFs~\cite{ABKM} are 
also given in Table~\ref{tab:cs}, for comparison. 
In both cases we use the PDFs in the 5-flavor scheme, which are 
derived from the 3-flavor ones (cf. also Ref.~\cite{JR})
using the matching conditions of 
Ref.~\cite{BMSN}. The $W$-, $Z$-, and Higgs cross sections are calculated 
in the NNLO QCD approximation~\cite{WZ,HIGGS}. The $t\bar{t}$- cross sections 
are calculated in the NLO approximation with a partial account of 
NNLO corrections~\cite{ttbar}. The impact of the PDF update
on the $W$ and $Z$ production cross sections is significant, particularly at 
Tevatron. This difference can be traced back to the upward shift of the 
quark distributions at small $x$, due to shift of the HERA data. The 
Higgs and $t\bar{t}$- production cross sections are quite sensitive to the 
value of $\alpha_{\rm s}$. Therefore the variation of these cross sections is 
defined by the balance in $\alpha_{\rm s}$ and the gluon 
distribution changes.
\begin{table}[h]
\begin{center}
\begin{tabular}{|c|c|c|c|c|}
\hline
   & $W^\pm$ (nb) & Z (nb) & $t\bar{t}$ (pb)& $H$ (pb)\\
   &  & & $(M_t=173~{\rm GeV})$ & $(M_H=150~{\rm GeV})$ \\
\hline
\hline
\multicolumn{5}{|c|}{Tevatron} \\ \hline
ABKM09 (updt.) & $26.8\pm0.3$ & $7.88\pm0.07$ & $6.81\pm0.12$ & $0.35\pm0.02$ \\
\hline
ABKM09 & $26.2\pm0.3$ & $7.73\pm0.08$ & $7.00\pm0.18$ & $0.36\pm0.03$ \\
\hline
\hline
\multicolumn{5}{|c|}{LHC (7 TeV)} \\ \hline
ABKM09 (updt.) & $100.9\pm1.3$ & $29.3\pm0.4$ & $132.7\pm6.0$ & $9.4\pm0.2$ \\
\hline
ABKM09 & $98.8\pm1.5$ & $28.6\pm0.5$ & $133.8\pm7.7$ & $8.8\pm0.3$ \\
\hline
\end{tabular}
\end{center}

\caption{Selected standard candle and other hard process cross sections 
for the Tevatron collider and the LHC calculated employing 
our updated NNLO PDFs in comparison to those of ABKM09 PDFs, Ref.~\cite{ABKM}.}
\label{tab:cs}
\end{table}

{\bf Acknoledgments.}{ This work was supported in part by Helmholtz Alliance
``Physics at the Terascale'', DFG Sonderforschungsbereich Transregio 9,
Computergest\"utz Theoretische Teilchenphysik and the European Commission 
MRTN HEPTOOLS under Contract No. MRTN-CT-2006-035505.}


\begin{thebibliography}{99}

\bibitem{ABKM}
  S.~Alekhin, J.~Bl\"umlein, S.~Klein and S.~Moch,
  %``The 3-, 4-, and 5-flavor NNLO Parton from Deep-Inelastic-Scattering Data
  %and at Hadron Colliders,''
  Phys.\ Rev.\  D {\bf 81} (2010) 014032.
  %%CITATION = PHRVA,D81,014032;%%

\bibitem{H1ZEUS}
  C.~Adloff {\it et al.}  [H1 Collaboration],
  Eur.\ Phys.\ J.\  C {\bf 21} (2001) 33;
  S.~Chekanov {\it et al.}  [ZEUS Collaboration],
  Eur.\ Phys.\ J.\ C {\bf 21}, 443 (2001).

\bibitem{HERA}
  F.~D.~Aaron {\it et al.}  [H1 Collaboration and ZEUS Collaboration],
  %``Combined Measurement and QCD Analysis of the Inclusive ep Scattering Cross
  %Sections at HERA,''
  JHEP {\bf 1001}, 109 (2010).
  %%CITATION = JHEPA,1001,109;%%


\bibitem{JB}
  I.~Bierenbaum, J.~Bl\"umlein and S.~Klein,
  %``Mellin Moments of the {$O(\alpha_s^3$)} Heavy Flavor Contributions to
  %unpolarized Deep-Inelastic Scattering at $Q^2 \gg m^2$ and Anomalous
  %Dimensions,''
  Nucl. Phys. {\bf B820} (2009) 417.

\bibitem{ML}
  E.~Laenen and S.~O.~Moch,
  %``Soft gluon resummation for heavy quark electroproduction,''
  Phys.\ Rev.\  D {\bf 59} (1999) 034027

\bibitem{MITOV}
  G.~Corcella and A.~D.~Mitov,
  %``Soft-gluon resummation for heavy quark production in charged-current  deep
  %inelastic scattering,''
  Nucl.\ Phys.\  B {\bf 676} (2004) 346.

\bibitem{VOGT}
  A.~Vogt,
  %``Constraining the proton's gluon density by inclusive charm
  %electroproduction at HERA,''
  arXiv:hep-ph/9601352.

\bibitem{AM}
  S.~Alekhin and S.~Moch,
  %``Higher order QCD corrections to charged-lepton deep-inelastic scattering
  %and global fits of parton distributions,''
  Phys.\ Lett.\  B {\bf 672} (2009) 166.


\bibitem{JR}
  P.~Jimenez-Delgado and E.~Reya,
  %``Variable Flavor Number Parton Distributions and Weak Gauge and Higgs Boson
  %Production at Hadron Colliders at NNLO of QCD,''
  Phys.\ Rev.\  D {\bf 80} (2009) 114011.

\bibitem{BMSN}
  M.~Buza, Y.~Matiounine, J.~Smith and W.~L.~van Neerven,
  %``Charm electroproduction viewed in the variable-flavour number scheme
  %versus fixed-order perturbation theory,''
  Eur.\ Phys.\ J.\  C {\bf 1} (1998) 301;\\
  I.~Bierenbaum, J.~Bl\"umlein and S.~Klein,
  %``The Gluonic Operator Matrix Elements at O(\alpha_s^2) for DIS Heavy Flavor
  %Production,''
  Phys.\ Lett.\  B {\bf 672} (2009) 401.


\bibitem{WZ}
R.~Hamberg, W.~L.~van Neerven and T.~Matsuura,
Nucl.\ Phys.\ B {\bf 359} (1991) 343
[Erratum-ibid.\ B {\bf 644} (2002) 403].

\bibitem{HIGGS}
  R.~V.~Harlander and W.~B.~Kilgore,
  %``Next-to-next-to-leading order Higgs production at hadron colliders,''
  Phys.\ Rev.\ Lett.\  {\bf 88} (2002) 201801.

\bibitem{ttbar}
   M.~Aliev, H.~Lacker, U.~Langenfeld, S.~Moch, P.~Uwer and M.~Wiedermann,
   %``-- HATHOR -- HAdronic Top and Heavy quarks crOss section calculatoR,''
   arXiv:1007.1327 [hep-ph].
   %%CITATION = ARXIV:1007.1327;%%


\end{thebibliography}
\end{document}